\documentstyle[aps,preprint,epsf]{revtex}

\begin{document}
\setcounter{page}{0}
\def\footnoterule{\kern-3pt \hrule width\hsize \kern3pt}
\tighten
\title{WHEN IS A SEMICLASSICAL APPROXIMATION SELF-CONSISTENT?
}

\author{Suzhou Huang
\footnote{Email address: {\tt shuang@mitlns.mit.edu}}}

\address{Center for Theoretical Physics \\
Laboratory for Nuclear Science \\
and Department of Physics \\
Massachusetts Institute of Technology \\
Cambridge, Massachusetts 02139}

\date{hep-ph/9605461. {~~~} May 1996}
\maketitle

\thispagestyle{empty}

\begin{abstract}
A general condition for the self-consistency of a semiclassical
approximation to a given system is suggested. It is based on
the eigenvalue distribution of the relevant Hessian evaluated
at the streamline configurations (configurations that almost satisfy
the classical equations of motion). The semiclassical approximation 
is consistent when there exists a gap that separates small and large 
eigenvalues and the spreading among the small eigenvalues is much
smaller than the gap. The idea is illustrated in the case of
the double-well potential problem in quantum mechanics. The 
feasibility of the present idea to test instanton models
of QCD vacuum is also briefly discussed.
\end{abstract}

\vspace*{\fill}
\begin{center}
Submitted to: {\it Physical Review D}
\end{center}

\section{Introduction}

Identification of the most important degrees of freedom in a complicated
physics process is often desirable and instructive. This identification
is particularly simple when a scale separation exists in the problem
under consideration. The energy gap, which divides the degrees of freedom
into light and heavy (or sometimes called slow and fast) modes, is a
necessary condition for such an endeavor to be successful.

The semiclassical approximation is one of the widely used technique in
achieving the isolation of the relevant degrees of freedom. Here we limit
the term semiclassical approximation to the expansion of functional
integrals around classical saddle point (or generically called instantons).
A natural question coming to mind is how the scale separation manifests
in this type of approximations. This question is easily answered in the
classical limit. In the context of the saddle point expansion the
relevant scale is set by the eigenvalues of the
Hessian evaluated at the saddle point. The Hessian here is defined
as the second derivative of the action with respect to the functional
integration variable evaluated at the saddle: 
$S''[\phi_{\text{saddle}}](x,y)
\equiv\delta^2 S[\phi_{\text{saddle}}]/\delta\phi(x)\delta\phi(y)$,
where $\phi_{\text{saddle}}$ satisfies the classical equation of motion
$\delta S[\phi_{\text{saddle}}]/\delta\phi(x)=0$. 

For theories without scale invariance, such as the one dimensional
double-well potential problem, the light modes are the zero modes arising
for symmetry reasons. The other non-vanishing eigenvalues, separated by some
intrinsic scale of the theory from the zero modes, correspond to the heavy
modes. As long as the coupling remains small, which in turn ensures the
diluteness of the instanton ensemble, this picture is generally preserved.
Under this circumstance identifying the collective motion of the nearly
zero modes as the relevant degrees of freedom is obvious.

For other theories, such as the two dimensional $O(3)$ non-linear
$\sigma$-model and the four dimensional $SU(N)$ Yang-Mills theory,
it is also possible to find modified Hessians such that the eigenvalue
spectra evaluated at isolated instanton backgrounds have gaps that
separate the zero modes from non-zero modes. The modified Hessians are
actually defined by using non-trivial space-time integration measures;
in general these measures transform the continuous spectra of the naive 
Hessians into discrete spectra by compactifying the space-time volume.
In the two dimensional $O(3)$ non-linear $\sigma$-model, the modified
Hessian shares the same eigenvalue spectrum with the naive Hessian for
an isolated instanton \cite{jevicki}. In contrast, the spectrum of the 
modified Hessian in the four dimensional $SU(N)$ Yang-Mills theory becomes
discrete, whereas the naive Hessian has a continuous spectrum in the
dilute limit \cite{thooft}~\footnote{I thank Ian Balitsky for reminding
me the latter fact.}. In the remaining part of this paper the word
Hessian always refers to the one whose eigenvalue spectrum is discrete
and has an explicit gap in a single instanton background.

However, the scale invariance of these theories can make the
eigenvalue gap arbitrarily small even in the best cases. In fact
the eigenvalues are proportional to $\rho^{-2}$, where $\rho$ is
the size parameter of the instanton. The size parameter in a scale
invariant theory, as well as the diluteness, cannot be controlled
externally, but are determined only by the specific dynamics.
Therefore, it is not obvious that the collective motions of the nearly
zero modes are the dominant configurations. In light of the lack of
a systematic and analytic method in analyzing these situations, we
have no alternative but to resort to numerical approaches.

In this paper we propose to explicitly examine the eigenvalue distribution
of the Hessian evaluated at the so-called streamline configurations
numerically. The concept of streamlines was originally introduced in
the context of semiclassical approximation in \cite{stream}. The precise
definition of the streamline in our numerical work is not too important.
Roughly speaking, streamlines are smooth configurations that almost satisfy
the classical equation of motion; the corresponding Hessians possess
parametrically small or even negative eigenvalues, which would be zero
modes in the dilute limit. As pointed out in \cite{stream}, although
they are generally not zero modes, the streamlines can not be treated
as Gaussian modes in the functional integration, but rather should be
regarded as part of the collective modes.

In practice, we obtain the streamlines from the thermalized
configurations by locally minimizing the action or cooling. Therefore,
a quantitative understanding of cooling is essential. In section II
we present a linearized theory of cooling based on the eigenvalue
structure of the Hessian. Then, we argue that the consistency of
the semiclassical approximation relies on the fact that there exists
a window in cooling time where the eigenvalue distribution of the
corresponding Hessian has the properties: i) there is a gap that
divides all eigenvalues into small and large ones; and ii) the
spreading among the small eigenvalues is small relative to the gap.
{\em Correlating} the information of the eigenvalue distribution with
the monitoring of the physical observable of interest as a function
of cooling time will enable us to unambiguously identify whether the
instanton ensemble is relevant to the particular physical observable
under consideration. As we will see, condition i) is to guarantee
that the large eigenmodes can be integrated out perturbatively; and
condition ii) is to guarantee that cooling does not distort too much
the small eigenmodes. In section III, we explicitly demonstrate
this idea in the one-dimensional quantum-mechanical problem of
the double-well potential.

Of course, any configuration will reach a dilute instanton plateau if
cooled long enough, independently of the initial condition. This is why
we have to correlate the information of the eigenvalue distribution
with the monitoring of the physical observables of interest. Only when
the measurement of the interested physical observables is insensitive
to the cooling within the cooling window mentioned above the information
from the eigenvalue distribution is truly relevant. Physically, the
conditions on the Hessian eigenvalues only guarantee that it is possible
to separate slow and fast modes (by cooling for instance), but we also
need to make sure that the specific observable be dominated by these
slow modes.

It is also important to keep in mind that the streamline configurations
should not be regarded as some kind of approximation to the original
thermalized configurations. Instead, the streamline configuration is
only used as a point in function space at which the semiclassical
expansion is performed, or about which the full theory can be possibly
linearized. For example, it would not be appropriate to assume in general
that the topological charge susceptibility measured using the streamline
configurations be the true topological charge susceptibility.

Finally, since our ultimate interest is to test various instanton
models of QCD vacuum, we will briefly discuss the feasibility of the
method proposed here in section IV. In addition, we also speculate
on how the sizes of the gap and the spreading of the would-be zero
modes are related to the average instanton size and inter-instanton
distance used in QCD phenomenology.

Even though cooling is used to obtain the streamline configurations,
there is nothing special or unique about this technique. In fact, the 
precise location of the streamlines in functional space is not too 
restrictive. We show in the Appendix that almost identical streamlines 
can be obtained by using the neural network technique, often employed in 
the usual image processing, in the quantum mechanics example with
a double-well potential.

\section{A linearized theory of cooling}

Since the streamline configurations are obtained from the thermalized
ones by using cooling, it is crucial to understand, at least at a
semi-quantitative level, what cooling is actually doing to configurations.
For simplicity we will use the continuum notations in this section,
because the concept is also valid in the continuum. The generalization
to lattice formulation is straightforward.

The streamline configurations are obtained from the thermalized
configurations by locally minimizing the action or cooling with the
relaxation equation, following the direction of the classical force,
\begin{equation}
{\partial \phi_t(x)\over\partial t}=-
{\delta S[\phi_t]\over \delta\phi_t(x)}\, ,
\label{relax}
\end{equation}
where the subscript $t$ labels the cooling time and the initial
configuration $\phi_{t=0}(x)$ is already in thermal equilibrium.
One can easily recognize that the above equation is the Langevin
equation \cite{langevin} with the white noise switched off. After
certain period of cooling the configuration becomes smooth and almost
satisfies the classical equation of motion. At this point (say at
$t=t_0$) the force term, $\delta S/\delta\phi_t$, is small and hence
the configuration can be regarded as a streamline, which we call
$\xi_{t_0}(x)$.

In general, Eq.(\ref{relax}) is difficult to solve, though sometimes
numerically tractable. In order to have an analytic understanding
let us linearize the force term near the streamline, i.e.
$\delta S/\delta\phi_t\propto\phi_t(x)-\xi_{t_0}(x)$. This
approximation is of course equivalent to approximating the action as
\begin{equation}
S[\phi_t]\approx S[\xi_{t_0}]+
\langle\!\langle\,(\phi_t-\xi_{t_0}),\,S'[\xi_{t_0}]\,\rangle\!\rangle
+{1\over 2}\langle\!\langle\,(\phi_t-\xi_{t_0}),
\,S''[\xi_{t_0}]\,(\phi_t-\xi_{t_0})\,\rangle\!\rangle\, ,
\label{S_2}
\end{equation}
where $\langle\!\langle\cdots\rangle\!\rangle$ denotes the scalar
product in $x$-space. How good is this linear approximation in
fact strongly correlates with how well the system can be treated
semiclassically.

The linearized Eq.(\ref{relax}) can be trivially solved by finding
the eigenvalues and eigenvectors of the Hessian $S''[\xi_{t_0}]$,
\begin{equation}
\sum_y S''[\xi_{t_0}](x,y)\,\psi_l(y)=\lambda_l\,\psi_l(x)\, .
\end{equation}
The eigenvalue label $l$ can be either discrete or continuous in general.
Expanding $\phi_t(x)-\xi_{t_0}(x)$ and $S'[\xi_{t_0}](x)$ in terms
of the complete basis $\{\lambda_l,\psi_l(x)\}$,
\begin{equation}
\phi_t(x)-\xi_{t_0}(x)=\sum_l c_l(t)\,\psi_l(x)
\quad\quad\text{and}\quad\quad
S'[\xi_{t_0}](x)=\sum_l f_l\,\psi_l(x)\, ,
\end{equation}
the linearized version of Eq.(\ref{relax}) can be immediately
integrated, yielding
\begin{mathletters}
\begin{equation}
c_l(t)=c_l(0)\, e^{-\lambda_l t}-{f_l\over\lambda_l}
\Bigl(1-e^{-\lambda_l t}\Bigr)\, .
\label{ctex}
\end{equation}
Since we are expanding around streamlines and hence $f_l$'s are generally
small, Eq.(\ref{ctex}) would be always dominated by the first term unless
when $|\lambda_l|\to 0$. In the latter case, Eq.(\ref{ctex}) becomes
\begin{equation}
c_l(t)=c_l(0)-f_l\, t\quad\text{when}\quad |\lambda_l|\to 0\, .
\label{ctsmlt}
\end{equation}
\end{mathletters}
Therefore, the eigen modes can be classified according to the magnitudes
of their eigenvalues. When the eigen mode obeys Eq.(\ref{ctex}) with the
first term being dominating (or in the subspace $|\lambda_l|\gg |f_l|$),
the corresponding coefficient of $\phi_l(x)$ is being damped (or magnified
when $\lambda_l<0$) by the factor $\exp(-\lambda_l t)$. On the other hand,
when the eigen mode obeys Eq.(\ref{ctsmlt}) (or in the subspace
$|\lambda_l|\to 0$), the corresponding coefficient of $\phi_l(x)$ is being
drifted by the classical force $f_l$. It is natural to identify the
subspace of $|\lambda_l|\gg |f_l|$ as Gaussian fluctuations and the
subspace of $|\lambda_l|\to 0$ as collective motions.

Now we explicitly see why we need the existence of a gap in the eigenvalue
distribution and why we need the eigenvalue spreading near zero to be small
relative to the gap. The first condition guarantees that the Gaussian
fluctuation can be safely treated perturbatively. The second condition
guarantees that cooling can be used effectively in practice to filter
out the Gaussian fluctuation without distorting too much the collective
coordinates. The spreading among the small eigenvalues, in fact, serves
as a measure of the strength of the inter-instanton interaction.

It should be recognized that the eigen basis $\{\lambda_l,\psi_l(x)\}$
is $t_0$ dependent, through the dependence of the streamline solution
$\xi_{t_0}(x)$ in the definition of the Hessian $S''[\xi_{t_0}]$.
However, this dependence is very mild if those two conditions in the
eigenvalue distribution are satisfied. In other words, there should
be a window in $t_0$ such that the eigen basis $\{\lambda_l,\psi_l(x)\}$
is rigid and almost independent of the precise location of $t_0$. One
of course in principle can expand $\phi_t(x)$ around any complete set.
The specialty about the streamline $\xi_{t_0}(x)$ is that the
quadratic approximation Eq.(\ref{S_2}) is more likely to work than
an arbitrary set, in the sense that $f_l$'s and $c_l(0)$'s are small,
and hence the drifting forces and anharmonic terms are ignorable.

Due to the interaction between instantons (even at the
classical level), the force term never exactly vanishes and some of
the eigenvalues can be slightly negative. Therefore, the streamline
configurations are only metastable with respect to cooling. To what
extent cooling can isolate a streamline configuration in fact also
crucially depends on whether the gap in the eigenvalue distribution
of the Hessian exists and on whether the eigenvalue spreading near zero
is smaller than the gap. This seemingly disadvantageous property actually
provides us a nice diagnosis for the self-consistency. 

Finally, it should be emphasized that eigenvalue distribution of the
Hessian is actually intrinsic, not necessarily only pertinent to the
interpretation of cooling. For example, conditions i) and ii), when
satisfied, will be reflected from the Monte Carlo dynamics (at least
for local updatings) by showing two characteristic auto-correlation
time scales, given by the inverses of the gap and the spreading.

\section{The double-well potential problem}
In this section we explicitly illustrate the idea outlined earlier
in the problem of the one dimensional double-well potential in
quantum mechanics, defined by the Hamiltonian
\begin{equation}
\hat{H}=-{1\over 2}{d^2\over d\phi^2}+V(\phi)\quad\quad\text{with}
\quad\quad V(\phi)={g^2\over 2}\Bigl(\phi^2-{1\over 4g^2}\Bigr)^2\, .
\end{equation}
In terms of the path integral formulation the corresponding Euclidean
action in a finite box of length $L$ is given by
\begin{equation}
S[\phi]=\int_0^L dx\,\biggl\{{1\over 2}\Bigl({d\phi(x)\over dx}\Bigr)^2
+{g^2\over 2}\Bigl(\phi^2(x)-{1\over 4g^2}\Bigr)^2\biggr\}\, .
\end{equation}
The energy level splitting between the first excited state and the
ground state as a function of the coupling constant,
$\Delta E(g)\equiv E_1(g)-E_0(g)$, is the primary concern here.
The nice feature of this problem is that instanton density can be
systematically controlled by dialing the coupling constant. Hence
we know whether the semiclassical picture is good or bad at a given
coupling.

\subsection{Known results}
The physics associated with $\Delta E(g)$ is well known \cite{book}.
In the weak coupling limit ($g$ small) $\Delta E(g)$ is dominated
by the dilute instanton configurations or the tunneling effect.
In this regime the semiclassical approximation is expected to be
good. As the coupling $g$ increases, the instanton density also
increases. The interaction between near-by instantons and other
long wavelength objects, such as the flucton introduced in 
\cite{flucton}, become important. Then we expect that the semiclassical
approximation fails in the strong coupling regime. We would like to
verify that the eigenvalue distribution of the Hessian evaluated in the
streamline configurations can tell us for what values of $g$ the
semiclassical approximation is reliable.

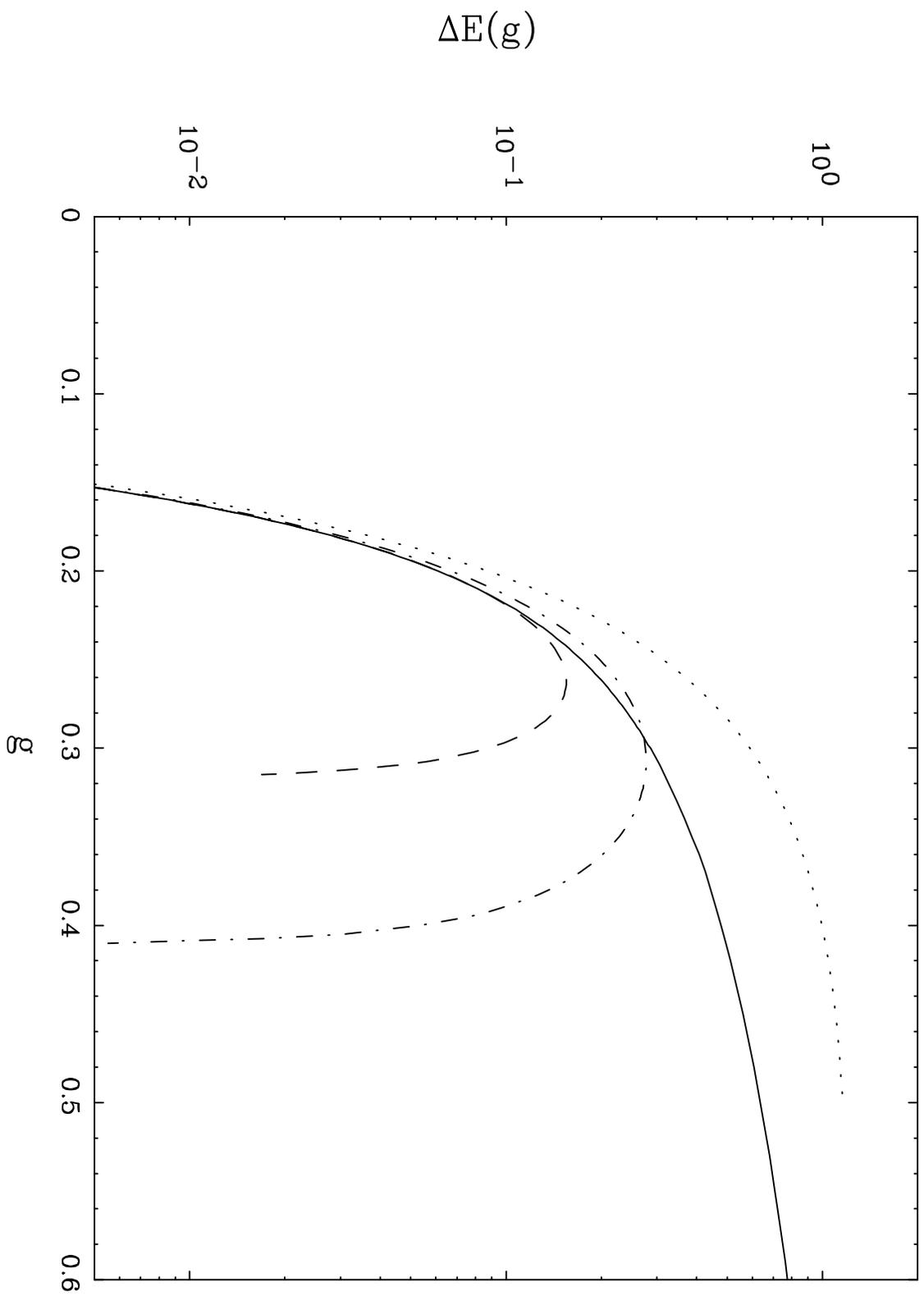
\begin{figure}
\caption{Energy gap between the first excited state and the ground
state as a function of the coupling $g$. The solid line is the exact
result. The dotted, dot-dashed and dashed lines are the one-loop, two-loop
and three-loop perturbative (around a single instanton) results.}
\label{fig1}
\end{figure}

To have a quantitative idea on where the semiclassical approximation
is valid we quote the perturbative (around one instanton) result
up to three loops \cite{three-loop,two-loop} for $\Delta E(g)$
\begin{equation}
\Delta E(g)={2\over\sqrt{\pi g^2}}\exp\Bigl(-{1\over 6g^2}\Bigr)\,
\biggl\{1-{71\over 12}g^2-{315\over 8}g^4+{\cal O}(g^6)\biggr\}\, ,
\end{equation}
which is plotted in Fig.\ref{fig1}. For comparison, the exact numerical
result, calculated using the method of moments recursion \cite{exact},
is also included. Because the perturbative expansion is divergent and 
non-Borel summable, the perturbative result quickly deteriorate at 
$g>0.25$. Therefore,
$g\sim 0.25$ can be regarded as a rough division between the weak and
strong coupling regimes. A similar value of $g$ can also be estimated
by requiring that the average distance between nearby instantons be
comparable with the instanton size (in the weak coupling limit the
instanton density $=\Delta E(g)$).

It is also useful to list the known properties of a well isolated
instanton in this theory. The continuum instanton solution is given by
$\phi_{\text{kink}}(x)={1\over 2g}\tanh({x\over 2})$ and the associated
action is $S_0=1/(6g^2)$. The eigenvalue problem of the Hessian in the
background of $\phi_{\text{kink}}(x)$ is equivalent to a one 
dimensional Schr\"{o}dinger equation with a potential 
$V''(\phi_{\text{kink}}(x))/2$
\begin{equation}
\biggl\{-{1\over 2}{d^2\over dx^2}+
\Bigl(3g^2\phi_{\text{kink}}^2(x)-{1\over 4}\Bigr)\biggr\}
\psi_l(x)=\lambda_l\,\psi_l(x)\, ,
\end{equation}
where $l$ labels eigenvalues. The solution of this equation has been
worked out long time ago \cite{spectrum}. There are two bound states
with $\lambda_0=0$, corresponding to the so-called translation zero
mode (with $\psi_0(x)\propto \phi'_{\text{kink}}$), and $\lambda_1=3/8$,
followed by a continuum with a threshold $\lambda_c=1/2$. These
features will be recognized later from the eigenvalue distribution
of the Hessian in the weak coupling regime. Since there is no scale
invariance in this model, the size of the instanton is fixed.

\subsection{Numerical results}
The thermalized configurations are generated by the standard Metropolis
method \cite{MC}, accompanied by the embedded cluster updating of
\cite{cluster}. Without the embedded cluster updating it would take
too long to have a good sampling of the action in the weak coupling
regime. For our purpose it is sufficient to use the simplest lattice
action
\begin{equation}
S_L[\phi]=a\, \sum_{n=1}^N \biggl\{
{1\over 2}\Bigl({\phi(n+1)-\phi(n)\over a}\Bigr)^2+
{g^2\over 2}\Bigl(\phi^2(n)-{1\over 4g^2}\Bigr)^2\biggr\}\, ,
\end{equation}
where $a$ is the lattice spacing and $N=L/a$. It is easy to verify that
the lattice action for a single instanton is related to its continuum
counterpart by $S_0^L/S_0=1+a^2/360+{\cal O}(a^4)$. Therefore, as long
as $a<1$ the discretization error is under control. In the simulation
$N$ has to be sufficiently large to ensure that enough instantons are
present in each thermalized configuration. As mentioned earlier, the
instanton density is equal to $\Delta E(g)$ in the weak coupling limit.
So we have chosen $L\,\Delta E(g)>10$, which translates into $N$ ranging
from 200 to 500 for the  various $g$'s we have considered. For such big 
values of $N$ the boundary effect is irrelevant, and we choose the periodic
boundary condition for simplicity.

In the Monte Carlo simulation, one iteration is defined as a Metropolis
sweep plus a embedded-cluster sweep. Typically, 500 iterations are used
to thermalize an arbitrary initial configuration. Then 1000 independent
configurations, separated by 50 iterations, are used in the measurement
in all cases considered.

Physical observables are measured as follows. The ground state energy
can be calculated using the Virial theorem 
\begin{equation}
E_0(g)=\Big\langle {1\over 2}\phi\,V'(\phi)+V(\phi)\Big\rangle\, .
\label{virial}
\end{equation}
The ground state wavefunction squared $\Psi^2_0(\phi)$ can be obtained
by histogramming the Monte Carlo history of $\phi(x)$ after
thermalization. The energy splitting between the ground state and the
first excited state can be read off from the exponential decay of the
two-point correlation function at large $|x|$
\begin{equation}
\Big\langle\phi(x)\phi(0)\Big\rangle\longrightarrow
\exp\Bigl(-\Delta E(g)|x|\Bigr)\, .
\label{tpf}
\end{equation}

The streamline configurations are obtained from the thermalized
configurations by locally minimizing the action iteratively or cooling,
\begin{equation}
\phi_{t+\epsilon}(n)=\phi_t(n)-\epsilon\,
\biggl\{{2\phi_t(n)-\phi_t(n+1)-\phi_t(n-1)\over a}+
2ag^2\phi_t(n)\Bigl[\phi_t^2(n)-{1\over 4g^2}\Bigr]\biggr\}\, ,
\label{dle}
\end{equation}
a discretized version of Eq.(\ref{relax}). The above iteration is
implemented in the serial mode (or parallel mode with checkerboard)
to avoid numerical instability. This implies that the cooling speed
does not exactly obey the analysis in the previous section, which holds
strictly in the parallel (without checkerboard) mode.
The instability in the parallel mode is due to the fact that those
modes with negative eigenvalues get magnified under Eq.(\ref{dle}).
A value of $\epsilon=0.1$ is found to be adequate in our
study. Violation of the classical equation of motion can be monitored
from the first derivative of the lattice action $S'_L[\phi_t]$
\begin{equation}
f[\phi_t(n)]\equiv {S'_L[\phi_t(n)]\over a}=
\biggl\{{2\phi_t(n)-\phi_t(n+1)-\phi_t(n-1)\over a^2}+
2g^2\phi_t(n)\Bigl[\phi_t^2(n)-{1\over 4g^2}\Bigr]\biggr\}\, .
\label{eom}
\end{equation}
The following mean value (relative to the location of the classical
vacua) is defined, which can serve as a quantitative measure
of the violation of the classical equation of motion on average
\begin{equation}
\bar{f}\equiv {2g\over N}\,\sum_{n=1}^N \Big| f[\phi_t(n)]\Big|\, .
\end{equation}

\subsubsection{An idealized case}
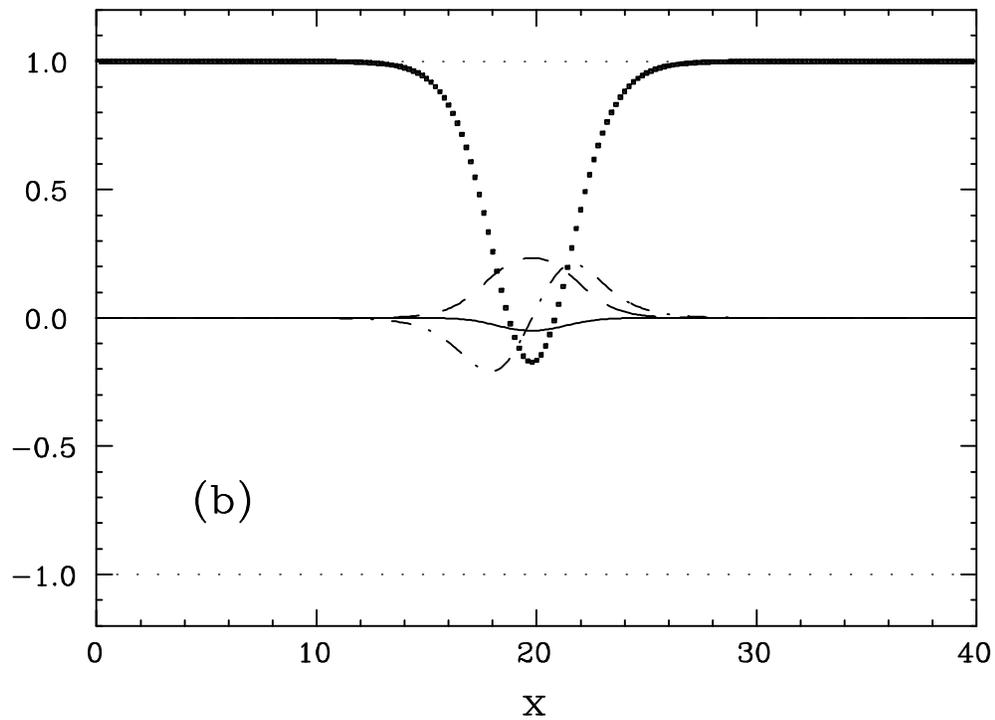
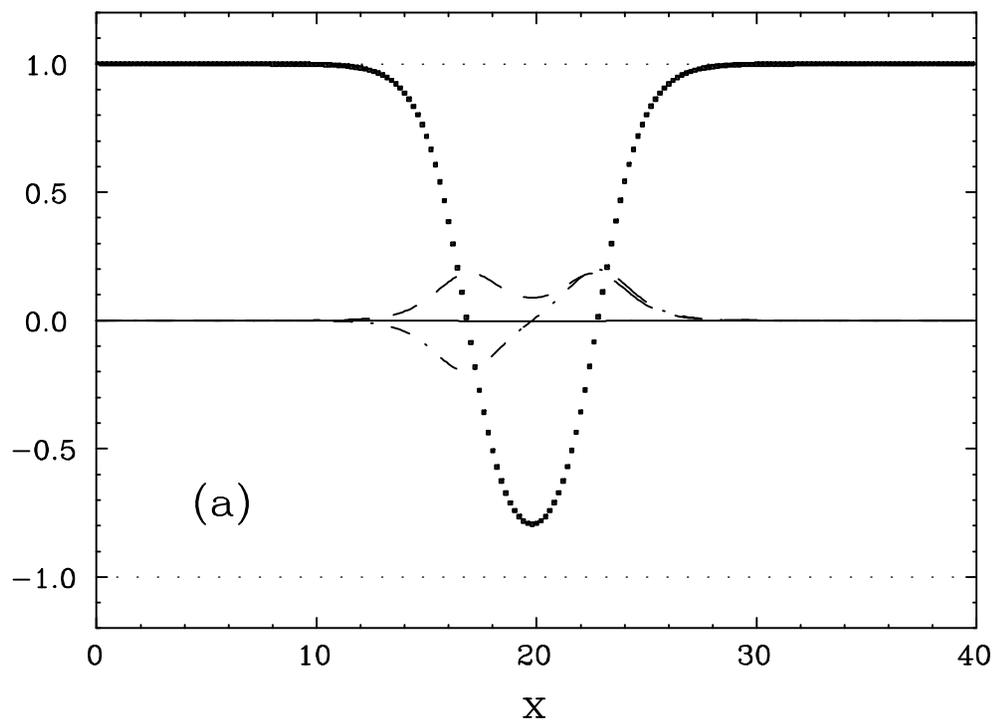
\begin{figure}
\caption{(a) Slightly overlapping instanton and anti-instanton and
(b) strongly overlapping instanton and anti-instanton pairs.
The heavy dots are $\phi_{\text{pair}}(x)$ at $g=0.5$ and
$a=0.2$. The solid line is the force term $S'[\phi_{\text{pair}}](x)$.
The dashed and dot-dashed lines are the lowest and second lowest
eigenvectors of $S''[\phi_{\text{pair}}](x,y)$ respectively. The thin
dotted lines indicate the two potential minima. }
\label{fig2}
\end{figure}

The streamline solution in an idealized case, a near-by instanton and
anti-instanton pair, has been studied some time ago in \cite{flucton}.
We add to that study by explicitly solving also the eigenvalues
and eigenvectors of the associated Hessian. In Fig.\ref{fig2}(a)
we show the streamline solution of a slightly overlapped instanton
and anti-instanton pair (heavy dots), the corresponding violation of the 
classical equation of motion (solid line) and the two eigenvectors with
the lowest eigenvalues, $\lambda^+=-0.025041$ (dashed line) and
$\lambda^-=0.004947$ (dash-dotted line), respectively.
Other eigenvalues are 0.308339 or higher. The separation
of the instanton and anti-instanton is roughly 6, while the ``radius''
of an instanton is about 3 (or a full size of about 6). The violation of
the classical equation of motion is hardly noticeable. If the instanton
and anti-instanton were widely separated the two lowest eigenvalues would
both be vanishing and the gap would be exactly $\lambda_1=3/8=0.375$.
Of course, a slight overlap lifts the degeneracy. The total action of
this streamline is 1.9683$S_0$.

Notice that one of the eigenvalue is still close to zero, corresponding
to the translation of the instanton and anti-instanton pair together.
The other eigenvalue becomes noticeably negative, due to the attraction
between the instanton and anti-instanton, corresponding to the relative
motion between the two objects. The minus sign of the lowest eigenvalue
implies that the system is not stable under cooling and the instanton
and anti-instanton eventually annihilate each other. However,
because $|\lambda^+|$ is much smaller than the gap, it is possible to
find a cooling-time window such that the higher modes are strongly 
damped while the pair is still almost intact. In the actual configuration
the spread of the small eigenvalues relative to the gap is a very good
indicator whether cooling is capable of faithfully separating collective
modes from Gaussian modes, or more precisely whether these two sets of
modes can be defined in a meaningful way.

When instanton and anti-instanton are closer, the violation of classical
equation of motion gets larger and the eigenvalue structure become very
different from the dilute limit. In Fig.\ref{fig2}(b), a pair with
separation comparable to the radius of instanton is displayed. Now the
lowest four eigenvalues are $-0.137194$, 0.116332, 0.379692 and 0.501244.
In this case, the gap is about only 3 times larger than the spreading
of the two small eigenvalues, and the total action is 1.4628$S_0$.
In this case the trace of instanton and anti-instanton is marginally
identifiable.

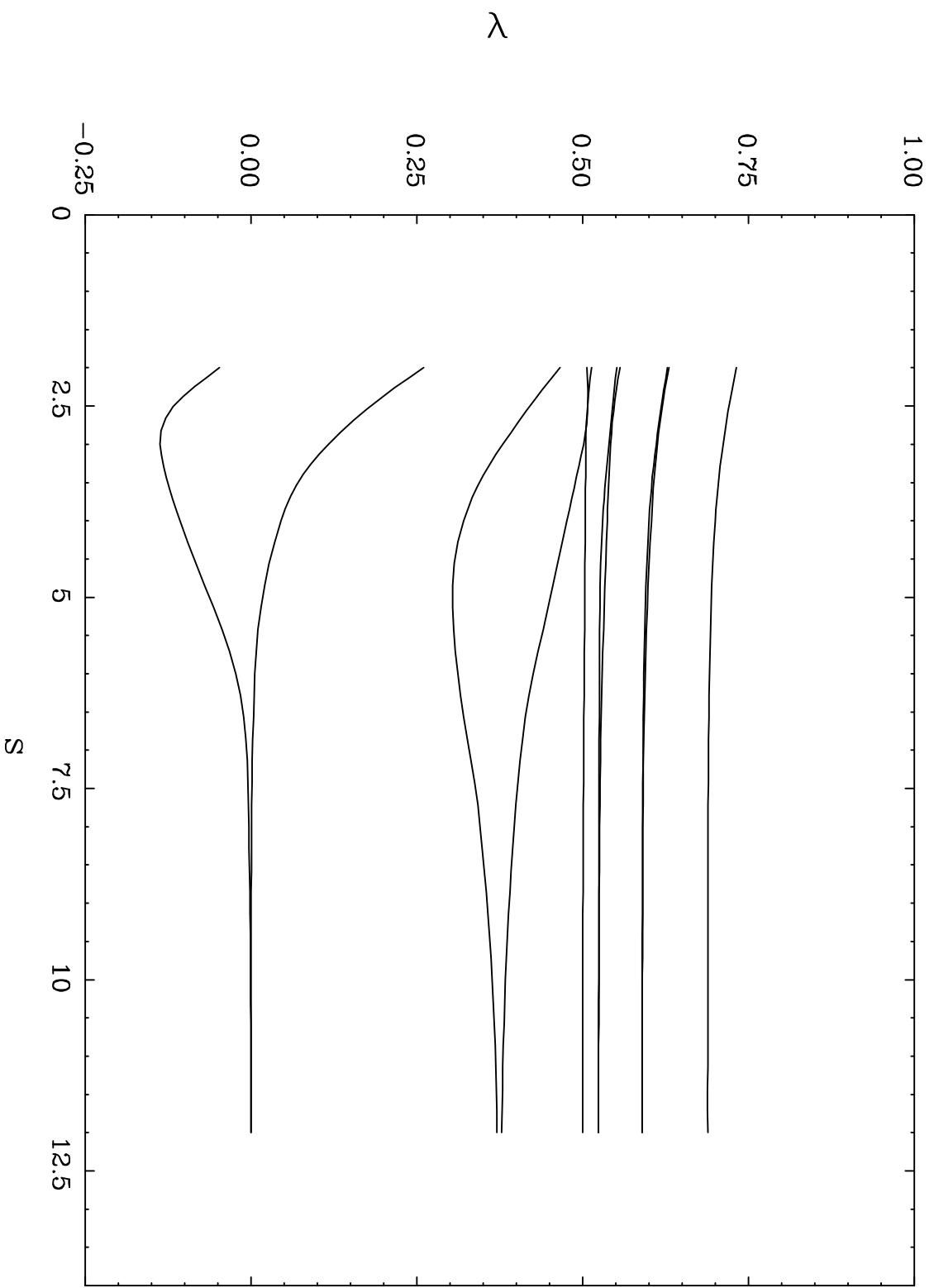
\begin{figure}
\caption{The lowest 10 eigenvalues of the Hessian as a function of the
instanton and anti-instanton separation. When the separation is comparable
with the instanton size ($\sim 6$) the eigenvalue pattern starts to deviate
from the dilute limit.}
\label{fig3}
\end{figure}

In Fig.\ref{fig3} the lowest ten eigenvalues are plotted as functions
of the instanton and anti-instanton separation $s$. When $s$ is large
the pattern of the eigenvalues are verified as that of the weak coupling
limit. When $s$ becomes comparable with the instanton size the lowest
four eigenvalues start to deform, while the higher eigenvalues stay more
or less the same. When $s$ is smaller than the radius of the instanton
the eigenvalue pattern reduces to that of the zero background field or
the plane waves.

\subsubsection{A weak coupling case}
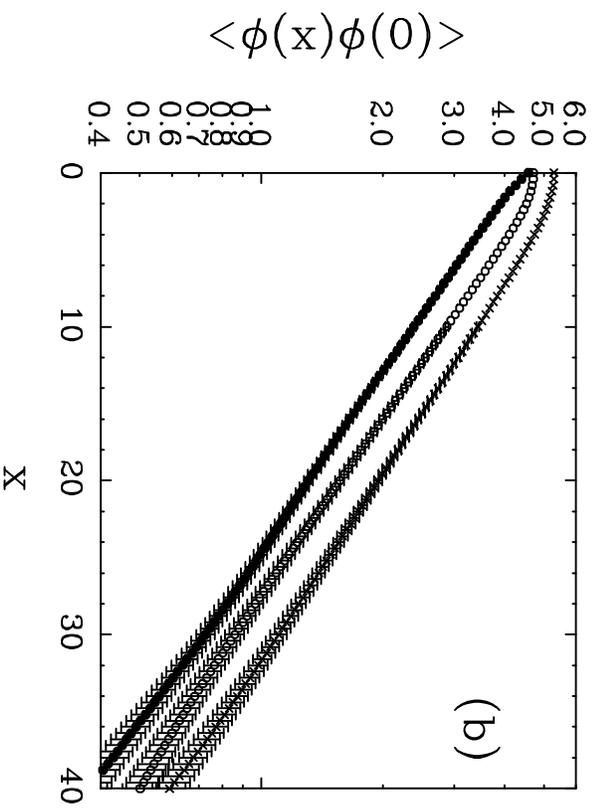
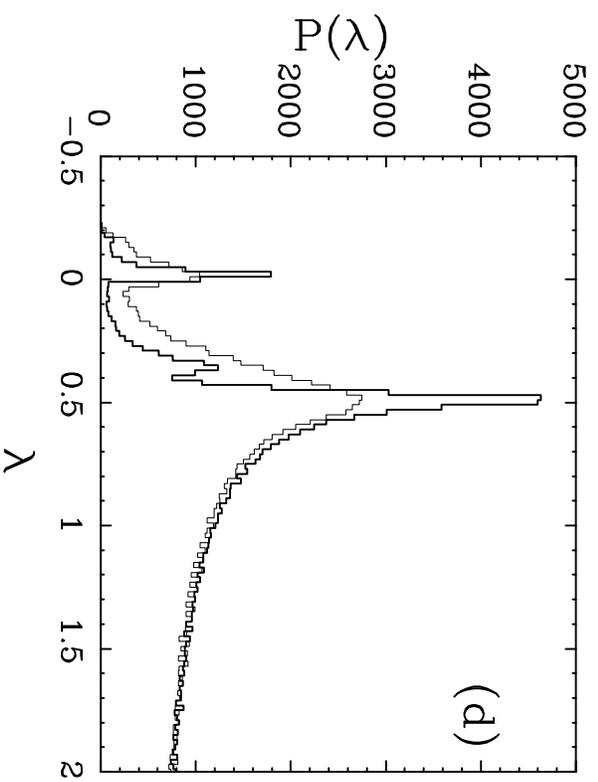
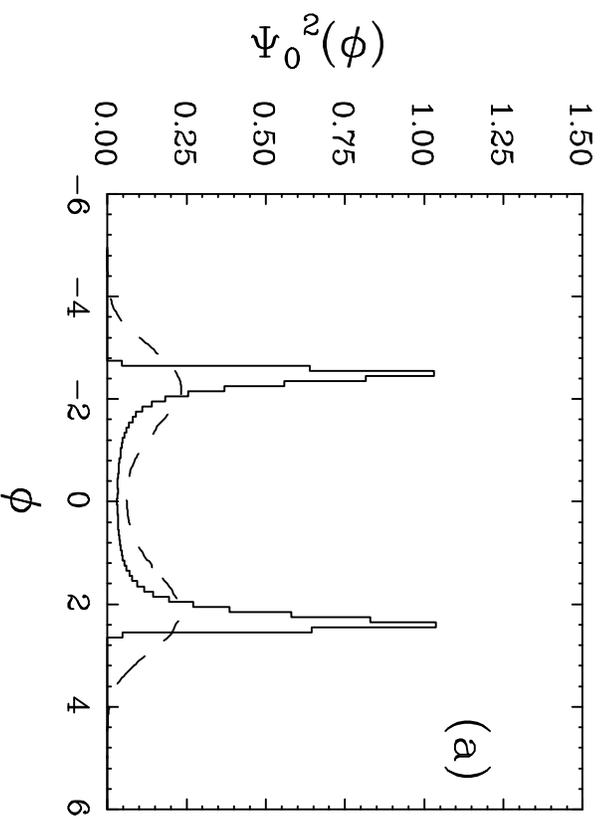
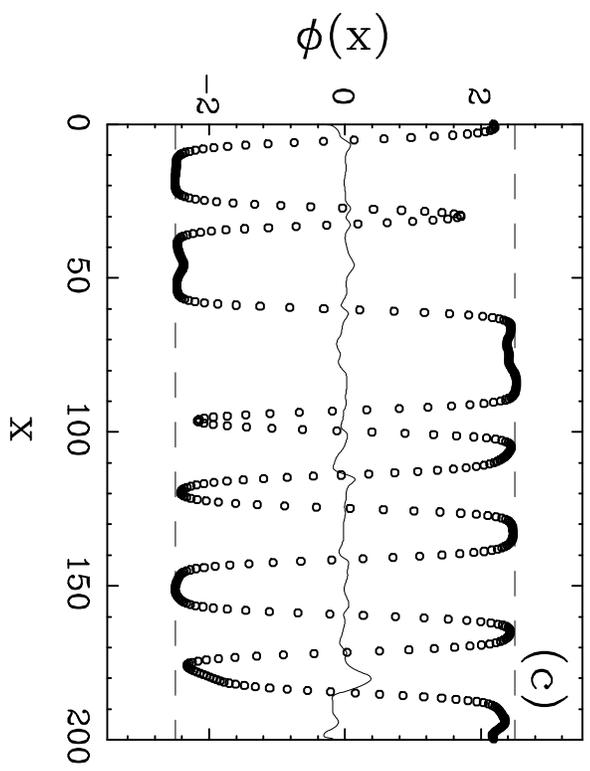
\begin{figure}
\caption{A weak coupling case: $g=0.20$, $a=0.4$ and $L=200$.
(a) ground state wavefunction squared $\Psi^2_0(\phi)$ at $N_{cool}=0$
(dashed line) and $N_{cool}=25$ (solid line); (b) two-point correlation
function $\langle\phi(x)\phi(0)\rangle$ at $N_{cool}=0$ (lower curve
of filled dots), $N_{cool}=25$ (middle curve of open dots) and 
$N_{cool}=50$ (upper curve of crosses). The errors are
purely statistical. When not shown the errors are smaller than symbols;
(c) a typical field configuration at $N_{cool}=50$ (open dots), the
violation of the classical equation of motion (thin solid line) and the
locations of the classical action minima (dashed lines); (d) eigenvalue
distribution of the Hessian evaluated at the streamlines $P(\lambda)$
at $N_{cool}=25$ (thin line) and $N_{cool}=50$ (thick line).}
\label{fig4}
\end{figure}

The weak coupling case we considered involves the following
parameters: $g=0.20$, $a=0.4$ and $L=200$ (or $N=500$).
Before cooling ($N_{cool}=0$) we measured the following quantities.
From Eq.(\ref{virial}) the ground state energy
$E_0=0.415(3)$, whose corresponding wavefunction squared is shown as
a dashed line in Fig.\ref{fig4}(a)). The two-point function is shown
in Fig.\ref{fig4}(b)) as dots (lower curve); from the values of the
two-point function  in the interval $x\in(6,30)$ and Eq.(\ref{tpf})
we extracted the energy splitting $\Delta E^{(0)}=0.060(2)$.
For comparison, the exact ground state energy and the energy gap at the 
same coupling are: $E_0^{(exact)}=0.4198$ and $\Delta E^{(exact)}=0.0609$.

A typical configuration at $N_{cool}=50$ is depicted by open dots in
Fig.\ref{fig4}(c). It is easily recognized that this configuration
is a dilute superposition of instantons and anti-instantons. The
violation of classical equation of motion is small ($\bar{f}=0.02442$),
as indicated by the thin line in Fig.\ref{fig4}(c).
Therefore, the configurations after $N_{cool}=50$ sweeps of cooling
are good candidates of streamlines.
At $N_{cool}=50$ we present the two-point function in Fig.\ref{fig4}(b)
with crosses (upper curve) and the eigenvalue distribution of the Hessian
with a thick-line histogram in Fig.\ref{fig4}(d). To make
sure that the cooling window indeed exists we also calculated the
two-point function (middle curve with open dots in Fig.\ref{fig4}(b)) 
and the eigenvalue distribution of the Hessian (thin line in 
Fig.\ref{fig4}(d)) at
$N_{cool}=25$. A comparison of the results at two different values of
cooling sweep gives us some idea on the evolution of these quantities
as functions of cooling time.

Since we are in the weak coupling regime, Fig.\ref{fig4}(d) clearly
show the anticipated properties: i) the existence of a gap that 
separates small and large eigenvalues and ii) the spreading among the 
small eigenvalues is small relative to the gap. Various peaks in
this figure can be easily identified. The peak near $\lambda=0$ is
associated with the ``would-be'' zero-modes, with an effective width
less than 0.1. The peak near $\lambda=0.35$ is associated with the
second discrete level mentioned in the last subsection. The third
peak represents the onset of the continuum at $\lambda=0.5$. The
effective gap can be roughly estimated to be between 0.35 to 0.5.

Fitting the cooled two-point functions with Eq.(\ref{tpf}) for $x$
in the interval $(6,30)$, we obtain $\Delta E^{(25)}=0.061(2)$
and $\Delta E^{(50)}=0.057(2)$ for $N_{cool}=25$ and $N_{cool}=50$,
respectively. These two numbers agree with the uncooled result
within statistical errors, as it is also visually seen in Fig.\ref{fig4}(b)
being the three curves nearly parallel. 
The measured ground state energy with these streamlines at $N_{cool}=25$
and $N_{cool}=50$ are very small, not surprisingly, since we do not
expect the ground state energy be dominated by the instanton physics. 
Correlating the informations from Fig.\ref{fig4}(b) and Fig.\ref{fig4}(d)
we can safely conclude that the energy splitting $\Delta E$ at $g=0.20$
is dominated by the dilute instanton physics, and hence can be
self-consistently treated semiclassically.

\subsubsection{A strong coupling case}
The strong coupling case we considered involves the following
parameters: $g=0.50$, $a=0.2$ and $L=40$ (or $N=200$).
Before cooling ($N_{cool}=0$) we measured the following quantities.
From Eq.(\ref{virial}) the ground state energy $E_0=0.293(5)$,
whose corresponding wavefunction squared is shown as
a dashed line in Fig.\ref{fig5}(a)). The two-point function is shown
in in Fig.\ref{fig5}(b)) as dots (lower curve); from the values of the
two-point function  in the interval $x\in(0.6,6)$ and Eq.(\ref{tpf})
we extracted the energy splitting $\Delta E^{(0)}=0.643(10)$.
For comparison, the exact ground state energy and the energy
splitting at the same coupling are: $E_0^{(exact)}=0.2940$ and
$\Delta E^{(exact)}=0.6374$.

Since we are in the strong coupling region there is no dilute
instanton physics associated with the energy splitting $\Delta E$ now.
We, therefore, anticipate that the information from monitoring
$\Delta E$ and the eigenvalue distribution of the Hessian as functions
of cooling time will indicate that the self-consistency of the
semiclassical approximation is violated. This expectation is borne
out explicitly.

The relevant results are shown in Fig.\ref{fig5}, with the same
notations as in Fig.\ref{fig4}. The energy splittings fitted from
data with $x\in(4,8)$ are $\Delta E^{(25)}=0.55(2)$ at $N_{cool}=25$,
and $\Delta E^{(50)}=0.43(2)$ at $N_{cool}=50$. These values
are much lower than the uncooled results, well beyond statistical errors, 
as can also be seen in Fig.\ref{fig5}(b) where the three sets of data 
points are no longer parallel. In addition, figure \ref{fig5}(d) shows
that the eigenvalue distribution does not have the two required 
properties i) and ii) even at $N_{cool}=50$. In fact,
the gap is clearly absent and the eigenvalue spreading
of the ``would-be'' zero modes is large. The absence of a gap and
the large spreading of the ``would-be'' zero modes can also be
inferred from a typical configuration after $N_{cool}=50$. In addition,
there are peaks showing up at large $\lambda$, whose position coincide
with the spectrum of the free Hessian ($-\nabla^2/2$):
$\lambda_n=[1-\cos(2\pi n/N)]/a^2$. 

From Fig.\ref{fig5}(c) we observe that the violation of the classical
equation of motion is large ($\bar{f}=0.09246$). Although the cooled
configuration is smooth, it is not a superposition of dilute instantons
and anti-instantons. This means that the theory cannot be linearized
around this kind of streamlines of poor quality. Even if one insists
that this kind of configurations be treated as streamlines, the force
terms would be so large that these streamlines would interact strongly,
hence a self-consistency could not be sustained. So we conclude that,
for $g>0.5$, there is no meaningful separation of collective modes
from Gaussian fluctuations. 

One could argue that the energy splitting can always be measured at
asymptotically large distances, even after many but fixed number of
cooling sweeps, since a finite number of cooling sweeps would not
modify correlations at distances much larger than the number of
cooling sweeps times the lattice constant.
However, the real relevant question is whether the
energy splitting can be extracted at a distances of the order of
$1/\Delta E$ right after the further excited states are damped out.
In the weak coupling case, we see that the window in $x$ used to extract
$\Delta E$ is very much the same before and after cooling. In contrast,
a similar window does not exist in the strong coupling case. The
cooled two-point functions do not even display clean exponential
behavior till $x>4$ or more.

\subsubsection{An intermediate coupling case}
The intermediate cooling case we considered involves the following
parameters: $g=0.35$, $a=0.2$ and $L=40$ (or $N=200$).
Before cooling ($N_{cool}=0$) we measured the following quantities.
From Eq.(\ref{virial}) the ground state energy $E_0=0.283(4)$,
whose corresponding wavefunction squared is shown as
a dashed line in Fig.\ref{fig6}(a)). The two-point function is shown
in in Fig.\ref{fig6}(b)) as dots (lower line); from the values of the
two-point function  in the interval $x\in(0.6,6)$ and Eq.(\ref{tpf})
we extracted the energy gap $\Delta E=0.399(7)$.
For comparison, the exact
ground state energy and the energy gap at the same coupling are:
$E_0^{(exact)}=0.2852$ and $\Delta E^{(exact)}=0.3870$.

Now we are in the intermediate situation. The results, depicted in
Fig.\ref{fig6}, lie somewhere between the weak and strong coupling
cases. The energy splitting fitted from data at $x\in(4,8)$ are
$\Delta E^{(25)}=0.36(2)$ ($N_{cool}=25$) and $\Delta E^{(50)}=0.33(2)$
($N_{cool}=50$); both values are slightly lower than the uncooled value. 
The violation of
the classical equation of motion after $N_{cool}=50$ is larger than
that of $g=0.20$, but smaller than that of $g=0.50$
($\bar{f}=0.05389$). It is interesting
to note how the eigenvalue distribution of the Hessian deforms as
a function of the coupling $g$. First, the peak associated with the
``would-be'' zero modes seems to be pushed to $\lambda\sim -0.15$.
Second, the strength of the peak at the threshold of the continuum
(at $\lambda=0.5$), together with the peak at the second bound state,
are now moved to the negative side. A similar
type of deformation can also be observed in Fig.\ref{fig5}(d). It
might not be easy to understand this kind of movement of eigenvalues
quantitatively. However, the observed deformation is certainly
qualitatively consistent with the attractive nature between instanton
and anti-instanton.

\section{Summary and conclusion}

We have presented a practical criterion for examining the
self-consistency of the semiclassical approach in approximating
functional integrals. It is based on the explicit eigenvalue
distribution of the Hessian evaluated in streamline configurations.
The self-consistency is guaranteed by two conditions: i) there exists
a gap that divides eigenvalues into small and large ones; and ii) the
spreading among the small eigenvalues is small relative to the gap.
We then illustrate how this idea can be explicitly applied in the
case of the one dimensional double-well potential problem in quantum
mechanics.

It should be noted that conditions i) and ii) do not implies that the
dynamics of streamlines is that of the dilute instanton limit. In
general, the dynamics of streamlines are most likely non-trivial.
Conditions i) and ii) only guarantees that the separation of the
collective motions and the Gaussian fluctuations can be made sensibly.

Of course, the ultimate goal is to apply the same method to QCD, and
to establish the self-consistency of the so-called instanton vacuum.
Technically, we anticipate no conceptual complications, apart that
the definition of the Hessian should be modified as mentioned in the 
Introduction and that the computation is obviously more demanding.
According to `t Hooft \cite{thooft} the modification of the Hessian
requires the knowledge of the location of instantons, which should not
be difficult to obtain once we have good candidates of streamline
configurations. In addition, the Hessian has similar structure to
the Dirac operator, apart from kinematics.
It was shown in \cite{thooft} that the modified Hessian
for gluons shares the same eigenvalue spectrum, except multiplicities,
as that of the modified Hessian for quarks in the continuum for a single
instanton. Since the eigenvalue problem of the lattice Dirac operator
is within the reach of today's computational resources, it may not
be totally unreasonable to assume that the method proposed here could 
be applied  to the Hessian associated with the gluonic action.

Furthermore, it is well known \cite{thooft} that the spectrum of the
modified Hessian of a well isolated instanton is discrete and the gap
is proportional to the inverse of the size parameter $\rho$ squared.
For realistic gluon configurations we expect this spectrum to be deformed.
To what extent the qualitative behavior of the eigenvalue distribution
is preserved will depend on two crucial scales: the average instanton
size ($\bar\rho$) and the inter-instanton distance ($\bar R$). It is
not difficult to imagine that, provided the instanton ensemble is not
too dense, $1/\bar\rho^2$ controls the size of the gap, and some
positive power of $1/\bar R$ controls the spreading of the would-be
zero modes.

In addition, phenomenologically, we roughly know that instantons
play a very important role in low energy light hadronic physics
\cite{vacuum}. This phenomenological picture, including the average
size of instantons ($\bar\rho\sim 0.3$ fm) and inter-instanton distance
($\bar R\sim 1.0$ fm), is confirmed to some extent by lattice calculations
\cite{cooling}. An explicit examination of the eigenvalue distribution of
the Hessian evaluated at streamline configurations in QCD will give us
direct and unambiguous information on how reliable the phenomenological
instanton models are.

\section{Acknowledgments}
I would like to thank Ian Balitsky, Richard Brower, Marcello Lissia,
John Negele and Kostas Orginos for many useful discussions. This work
is supported in part by funds provided by the U.S. Department of Energy
(D.O.E.) under cooperative research agreement \#DF-FC02-94ER40818.

\appendix
\section*{Streamline via Neural Network}
In the main text of this paper we used the cooling procedure to obtain
the streamline configuration. This is due to the convenience of physical
interpretation. In this appendix we show that almost identical streamline
can be obtained by using neural network technique, often employed in usual
image processing \cite{nn}.

The idea is to introduce $N$ linear units $\theta(n)$, whose output
represents the smoothened configuration at each lattice site. Then the
value of $\theta(n)$ is obtained by minimizing the following cost function
\begin{equation}
E_{\text{cost}}=\sum_{n=1}^N\,\biggl\{
\Bigl[\theta(n)-\phi_{t=0}(n)\Bigr]^2+
\alpha\Bigl[\theta(n)-\theta(n+1)\Bigr]^2\biggr\}\, .
\label{cost}
\end{equation}
The first term is to enforce the fidelity of the neural network output to
the original configuration $\phi_{t=0}$. The second term is a physically
motivated bias that constrains the neural network output to be smooth.
The relative importance of these two terms is controlled by the parameter
$\alpha$. Since the cost function $E_{\text{cost}}$ is quadratic in
$\theta$'s, the unique minimum is guaranteed. The output $\theta$'s at
the minimum, once $\alpha$ is properly chosen, can be regarded as a
candidate of the streamline configuration. The minimization can be
achieved by using, for example, the standard conjugate gradient method.

In Fig.\ref{fig7} we show a typical configuration (in dots) at $a=0.25$
and $g=0.25$. The standard cooling procedure with $\epsilon=0.1$ and
$N_{cool}=10$ yields the thick line, while the neural network smoothing
with $\alpha=10$ yields the thin line. The difference of the thick line
and the thin line is tiny. This result is easy to
anticipate, because the gradient descent of Eq.(\ref{cost}) with respect
to $\theta$ is very similar to the discrete Langevin equation,
Eq.(\ref{dle}).

In general, the value of $\alpha$ needs to be experimented, just like
the choice of $\epsilon$ and $N_{cool}$ in the cooling. Obviously,
various variations need to be incorporated in order to extend
Eq.(\ref{cost}) to other theories. More sophisticated neural
network techniques involving learning (supervised or unsupervised)
can also be imagined. Of course, the drawback of this approach is
that the precise physical content is theoretically less transparent.

\begin{figure}
\caption{A strong coupling case: $g=0.50$, $a=0.2$ and $L=40$.
Notations are the same as that of Fig. 4.}
\label{fig5}
\end{figure}

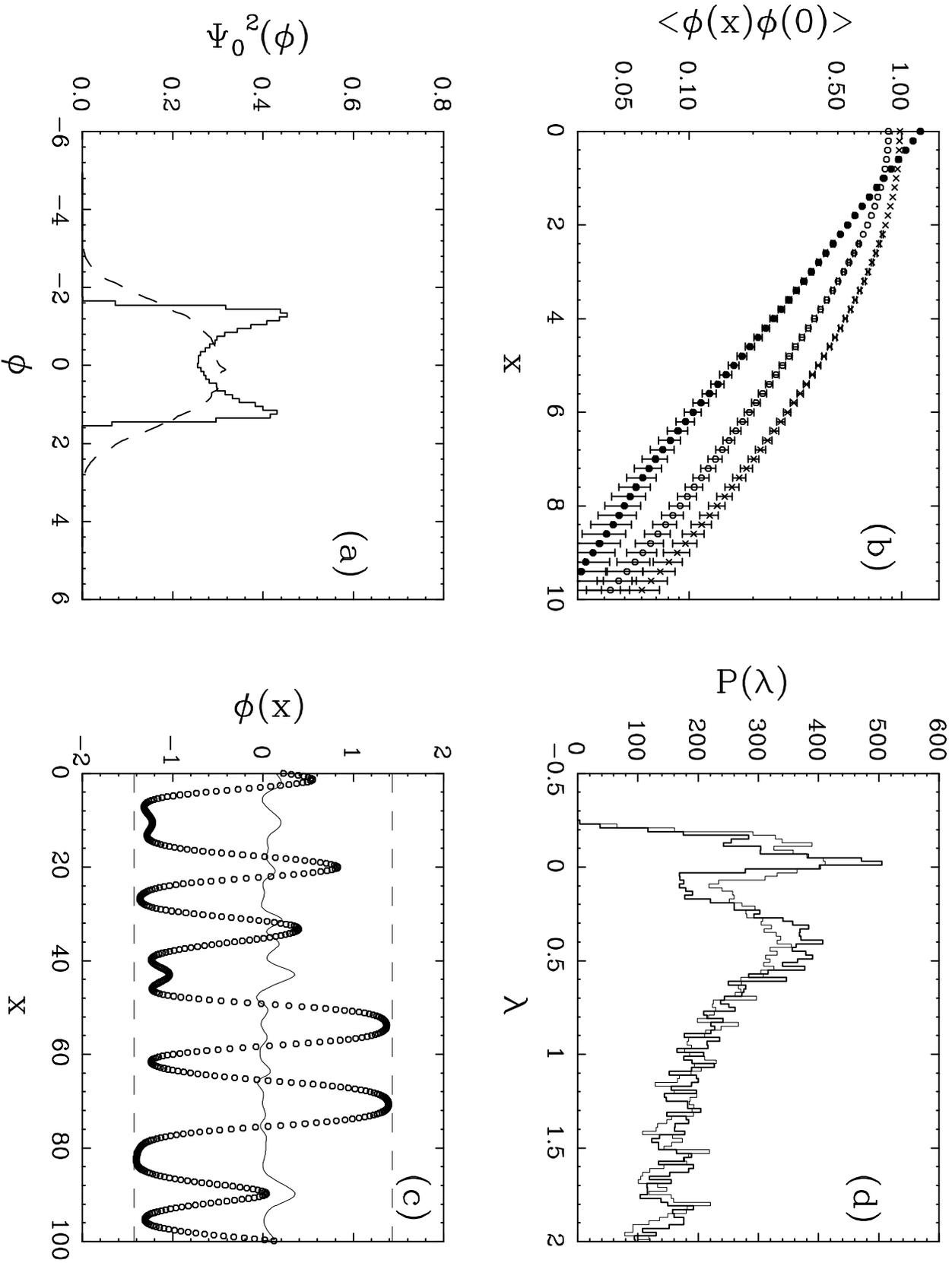
\begin{figure}
\caption{An intermediate coupling case: $g=035$, $a=0.2$ and $L=40$.
Notations are the same as that of Fig. 4.}
\label{fig6}
\end{figure}

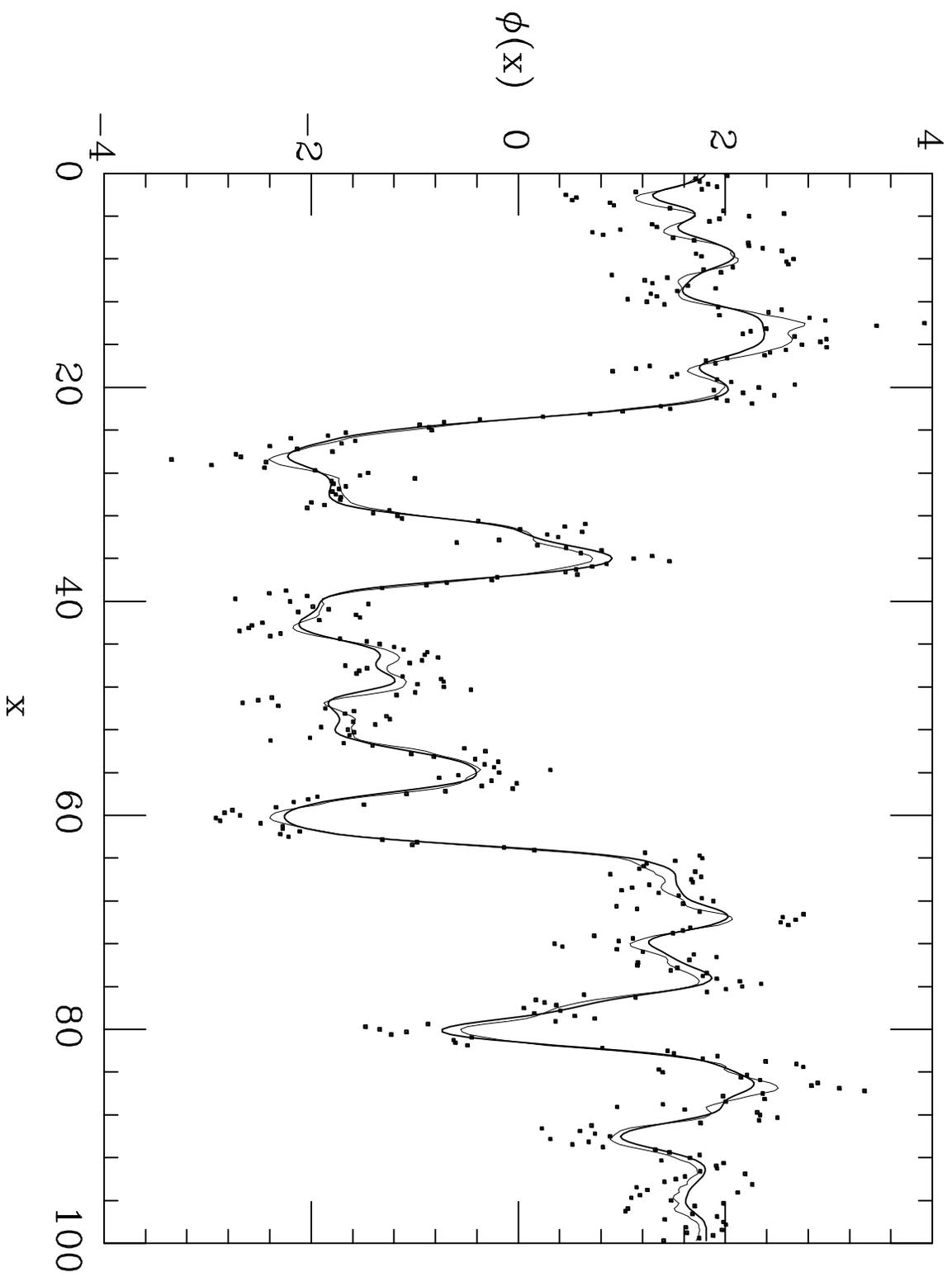
\begin{figure}
\caption{A typical configuration (in dots) at $a=0.25$, $g=0.25$ and
$L=100$. The standard cooling with $\epsilon=0.1$ and $N_{cool}=10$
yields the thick line. The neural network smoothing with $\alpha=10$
yields the thin line, which is almost indistinguishable from the
thick line.}
\label{fig7}
\end{figure}


\begin{references}

\bibitem{jevicki}
A.~Jevicki, Nucl. Phys. B127 (1977) 125;

D.~F\"{o}rster, Nucl. Phys. B130 (1977) 38.

\bibitem{thooft}
G.~'t~Hooft, Phys. Rev. D14 (1976) 3432.

\bibitem{stream}
I.~I.~Balitsky and A.~V.~Yung, Phys. Lett. B168 (1986) 113;

E.~V.~Shuryak, in ``Proc. Conf. on numerical experiments in quantum
field theories'', Alma-Ata, 1985, ed. A.~A.~Migdal (in Russian).

\bibitem{langevin}
G.~G.~Batrouni, et al., Phys. Rev D32 (1985) 2736.

\bibitem{book}
A very good discussion of the double-well potential problem from
the semiclassical point of view can be found in the book by
J.~Zinn-Justin, ``Quantum Field Theory and Critical Phenomena'',
2nd edition, Oxford Univ. Press, (1993).

\bibitem{flucton}
E.~V.~Shuryak, Nucl. Phys. B302 (1988) 621.

\bibitem{three-loop}
J.~Zinn-Justin, J.~Math. Phys. {\bf 22} (1981) 511;

\bibitem{two-loop}
C.~F.~W\"{o}hler and E.~Shuryak, Phys. Lett. B333 (1994) 467.

\bibitem{exact}
R.~Blankenbecler, T.~DeGrand and R.~L.Sugar, Phys. Rev. D21 (1980) 1055.

\bibitem{spectrum}
R.~F.~Dashen, B.~Hasslacher and A.~Neveu, Phys. Rev. D10 (1974) 4130.

\bibitem{MC}
M.~Creutz and B.~Freedman, Ann. of Phys. {\bf 132} (1981) 427;

E.~V.~Shuryak and O.~V.~Zhirov, Nucl. Phys. B242 (1984) 393;

\bibitem{cluster}
R.~C.~Brower and P.~Tamayo, Phys. Rev. Lett. {\bf 62} (1989) 1087.

\bibitem{vacuum}
For existing literature on the instanton models of the QCD vacuum, see
E.~V.~Shuryak, Rev. Mod. Phys. {\bf 65} (1993) 1; and a forthcoming
review article by T.~Shafer, E.~V.~Shuryak and J.~J.~M.~Verbaarschot.

\bibitem{cooling}
M.~Chu and S.~Huang, Phys. Rev. D45 (1992) 2446.

M.~Chu, J.~Grandy, S.~Huang and J.~Negele, Phys. Rev. D49 (1994) 6039.

\bibitem{nn} see for example, J.~Hertz, A.~Krogh and R.~G.~Palmer,
``Introduction to the Theory of Neural Computation'',
Addison-Wesley (1991), Reading, Mass.

\end{references}
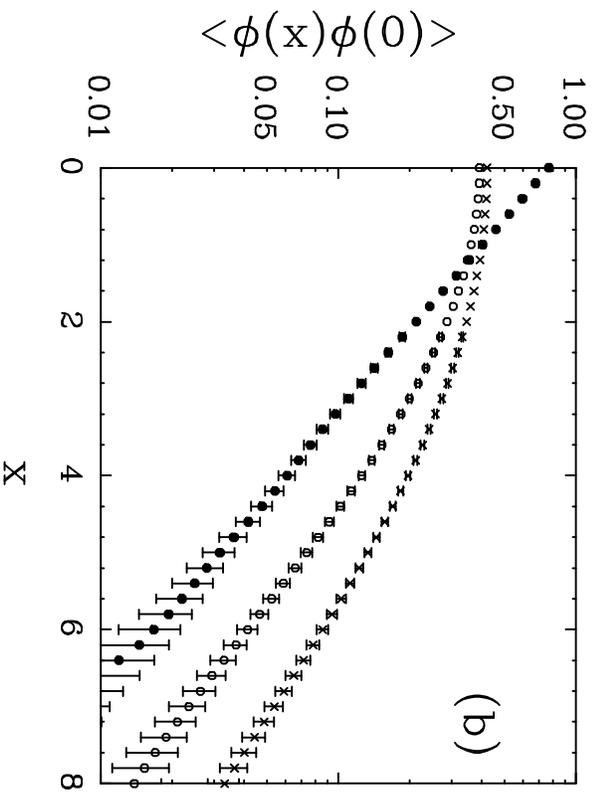
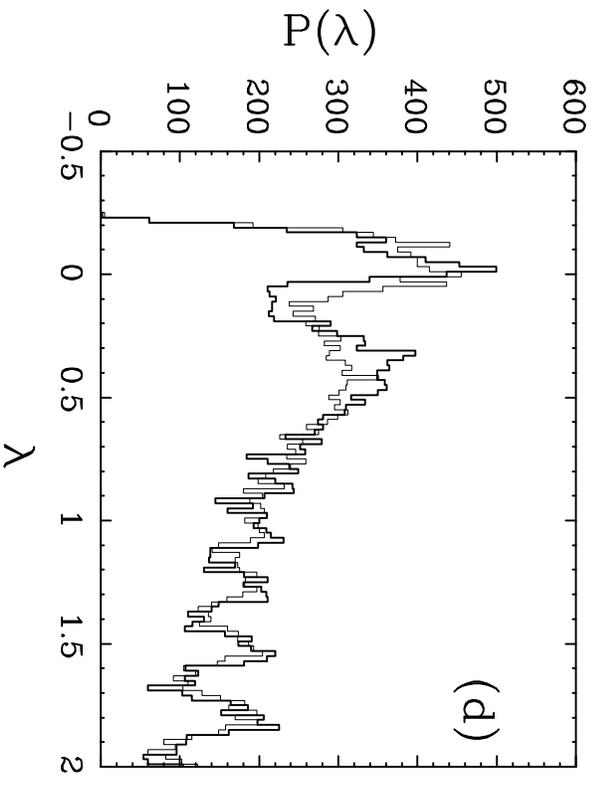
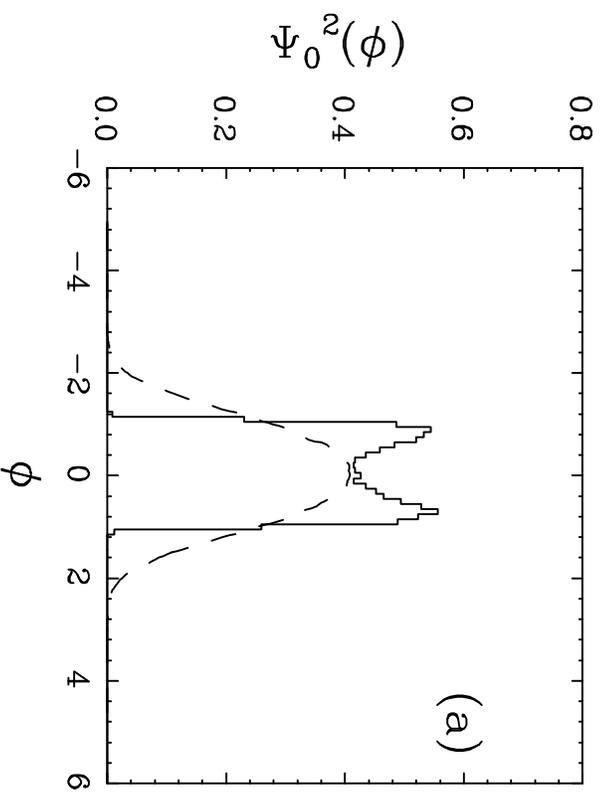
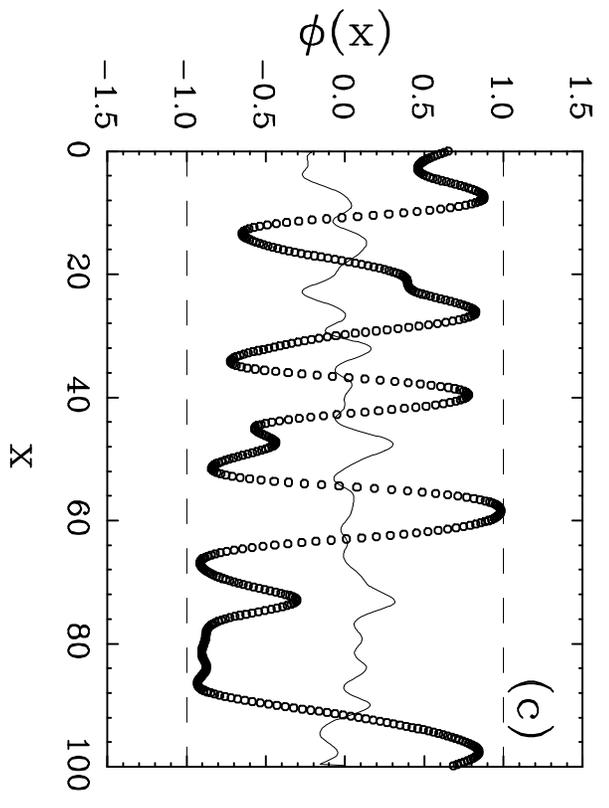
\end{document}